\documentclass[conference, a4paper]{IEEEtran}
\IEEEoverridecommandlockouts
\usepackage{amsmath,amssymb,bm,amsfonts,enumerate,url,footnote,color,ifpdf}
\usepackage[noend]{algpseudocode}
\ifCLASSINFOpdf
	\usepackage[pdftex]{graphicx}
         \usepackage{lipsum}
\else
	\usepackage[dvips]{graphicx}
         \usepackage{lipsum}
\fi
\usepackage{array,multirow,cite}
\ifCLASSOPTIONcompsoc
	\usepackage[caption=false,font=footnotesize,labelfont=sf,textfont=sf]{subfig}
\else
	\usepackage[caption=false,font=footnotesize]{subfig}
\fi
\usepackage[english]{babel}
\usepackage{booktabs,rotating,balance,arydshln} 
\usepackage{tikz} \usetikzlibrary{matrix}
\usepackage{algorithm}

\usepackage{titlesec}

\usepackage{booktabs} 
\usepackage{tabularx} 

\begin{document}
\title{Federated Deep Reinforcement Learning-Driven O-RAN for Automatic Multirobot Reconfiguration}
\author{
    \IEEEauthorblockN{Faisal Ahmed\IEEEauthorrefmark{1}, Myungjin Lee\IEEEauthorrefmark{2}, Shao-Yu Lien\IEEEauthorrefmark{4}, Suresh Subramaniam\IEEEauthorrefmark{3}, Motoharu Matsuura\IEEEauthorrefmark{7}, \\ Hiroshi Hasegawa\IEEEauthorrefmark{8}, and Shih-Chun Lin\IEEEauthorrefmark{1}} 
    \vspace{0.5em}
    \IEEEauthorblockA{\IEEEauthorrefmark{1}North Carolina State University, Raleigh, NC, USA;
     \{fahmed5, slin23\}@ncsu.edu} 
    \IEEEauthorblockA{\IEEEauthorrefmark{2}Cisco Research, Cisco Systems, Inc., CA, USA; 
    myungjle@cisco.com} 
    \IEEEauthorblockA{\IEEEauthorrefmark{4}National Yang Ming Chiao Tung University, Tainan, Taiwan; sylien@nycu.edu.tw}

    \IEEEauthorblockA{\IEEEauthorrefmark{3}The George Washington University, Washington, USA; 
    suresh@gwu.edu} 
    \IEEEauthorblockA{\IEEEauthorrefmark{7}University of Electro-Communications, Chofu, Japan; 
    m.matsuura@uec.ac.jp} 
    \IEEEauthorblockA{\IEEEauthorrefmark{8}Nagoya University, Nagoya, Japan; 
    hasegawa@nuee.nagoya-u.ac.jp}
    
}

    
\setlength{\baselineskip}{10.53pt} 
\maketitle
\thispagestyle{empty}

\begin{abstract}
The rapid evolution of Industry 4.0 has led to the emergence of smart factories, where multirobot system autonomously operates to enhance productivity, reduce operational costs, and improve system adaptability. However, maintaining reliable and efficient network operations in these dynamic and complex environments requires advanced automation mechanisms. This study presents a zero-touch network platform that integrates a hierarchical Open Radio Access Network (O-RAN) architecture, enabling the seamless incorporation of advanced machine learning algorithms and dynamic management of communication and computational resources, while ensuring uninterrupted connectivity with multirobot system. Leveraging this adaptability, the platform utilizes federated deep reinforcement learning (FedDRL) to enable distributed decision-making across multiple learning agents, facilitating the adaptive parameter reconfiguration of transmitters (i.e., multirobot system) to optimize long-term system throughput and transmission energy efficiency. Simulation results demonstrate that within the proposed O-RAN-enabled zero-touch network platform, FedDRL achieves a 12\% increase in system throughput, a 32\% improvement in normalized average transmission energy efficiency, and a 28\% reduction in average transmission energy consumption compared to baseline methods such as independent DRL.

\end{abstract}

\IEEEpeerreviewmaketitle

\section{Introduction}
Industry 4.0 represents a transformative shift in manufacturing, driven by the integration of advanced technologies such as artificial intelligence (AI), automation, and the Industrial Internet of Things (IIoT), leading to the emergence of smart factories\cite{9272626}. These factories are characterized by interconnected IIoT devices, such as \textbf{multirobot system (a group of robots that communicate and collaborate)} and industrial machines, operating autonomously to optimize productivity, reduce costs, and enhance real-time adaptability. However, managing the complexity of diverse industrial equipment and ever-changing network conditions poses substantial challenges. Consequently, sophisticated network automation solutions are critical to ensure reliable, low-latency communications and to dynamically handle resource allocation, and system optimization in real time \cite{9272626}.

Zero-touch networks have emerged as a key enabler of high-level automation in these industrial environments \cite{10004601}. By leveraging programmable interfaces and virtualization technologies (e.g., software-defined networking), zero-touch networks can proactively configure and optimize themselves with minimal or no human intervention \cite{9913206}. This end-to-end automation is especially vital in Industry 4.0 settings, where the diversity of IIoT devices, strict performance requirements, and rapid fluctuations in network demands render manual management impractical. Furthermore, the incorporation of machine learning (ML) into zero-touch networks allows data-driven decision-making that can accelerate response times and boost overall system efficiency \cite{10004601, 10004596}.

A significant advancement in zero-touch networking is the integration of the Open Radio Access Network (O-RAN) architecture \cite{o-ran2023, 9376232} which decouples network control functions from the underlying physical infrastructure \cite{10355063}. This design provides more flexible orchestration of communication and computational resources\cite{Ahmed2025}, accommodating the heterogeneous requirements found in smart factories. Key O-RAN components, such as the open central unit (O-CU), distributed unit (O-DU), and radio unit (O-RU) can serve as the backbone for advanced industrial use cases, such as the management of large-scale multirobot systems. By merging O-RAN with zero-touch networks, factory operators can automate parameter reconfiguration, such as modulation and coding scheme (MCS) selection and transmission power control for multirobot systems, thereby facilitating more effective adaptation to long-term operational optimizations in factory settings.

Federated learning (FL) \cite{FedAvg, 10138331} serves as a crucial adjunct to this advanced automation paradigm by enabling distributed model training without the need for centralizing data. In networked systems, FL has demonstrated its effectiveness in preserving data privacy \cite{FedAvg}, reducing communication overhead\cite{9631391}, and enhancing network performance \cite{Ahmed2025}. When incorporated into O-RAN-enabled zero-touch networks, FL can further support distributed intelligence for tasks such as resource allocation\cite{9771700}, and parameter reconfiguration. Despite these advantages, most existing studies that use FL in O-RAN environments \cite{Asad2024, 9771700} and zero-touch networks \cite{10130620, 9838376} either depend on traditional deep learning methods \cite{10012789} or do not fully exploit the hierarchical O-RAN architecture \cite{o-ran2023} or the potential of federated deep reinforcement learning (FedDRL) \cite{10416344} in heterogeneous industrial settings. Furthermore, the integration of FedDRL and hierarchical O-RAN architecture into zero-touch networks for Industry 4.0 remains largely unexplored in the existing literature, leaving a gap in effective solutions for dynamic MCS and transmission power adjustment.

To bridge these gaps, this paper introduces a novel hierarchical O-RAN-enabled zero-touch network platform tailored for smart factory scenarios, as illustrated in Figure 1. The introduced platform ensures seamless connectivity and efficient management of both the multirobot system and the underlying communication and computational infrastructure. A key feature of this platform is the integration of a FedDRL framework, which goes beyond conventional single-agent or centralized approaches to enable adaptive parameter reconfiguration of transmitters within the multirobot system. 
The main contributions are summarized as follows:

\begin{itemize}

\item To enhance long-term system throughput, improve transmission energy efficiency, and enable real-time adaptation to local conditions, the introduced hierarchical O-RAN-enabled zero-touch network platform integrates FedDRL. This incorporation facilitates intelligent parameter reconfiguration of transmitters by dynamically adjusting MCS selection and transmission power control, ensuring optimal performance in dynamic industrial environments.
\item The FedDRL framework is built on a Dueling Double Deep Q-Network (D3QN) architecture, incorporating prioritized experience replay (PER) to improve sample efficiency and momentum-optimized gradient descent (MGD) to accelerate convergence during local and global model parameter updates.

\item To ensure optimal transmission performance under diverse network conditions, a global reward function is designed that ensures the simultaneous maximization of system throughput and transmission energy efficiency. 

\item Extensive simulations have been conducted, and the results indicate that FedDRL provides substantial performance enhancements. Specifically, FedDRL demonstrates a 12\% and 32\% improvement in system throughput and normalized average transmission energy efficiency, respectively, compared to the baseline independent DRL (IDRL), and achieves a 28\% reduction in average transmission energy consumption relative to the baseline.
\end{itemize}

\section{Integration of O-RAN into the AI-Driven Zero-Touch Network Platform}
As illustrated in Figure 1, the O-RAN architecture is integrated into the proposed zero-touch network platform to enable next-generation networking systems. This integration facilitates dynamic and efficient radio optimization, flexible factory operations, and seamless management of transmitters by organizing field infrastructure into various hierarchical units within the architecture.

For \textit{communication infrastructure}, the E2 node in O-RAN comprises several network layer functions, including the O-CU, O-DU, and O-RU. The O-RU manages lower physical (PHY) layer functions, the O-DU handles medium access control (MAC), link control, and higher PHY layer functions, while the O-CU oversees radio resource control (RRC) and packet data convergence protocol (PDCP) for connection management, as shown in Figure 1. The O-CU can manage multiple O-DUs, each of which can oversee multiple O-RUs, ensuring fundamental connectivity within the O-RAN architecture. These units can be combined in various configurations to suit different scenarios. In our scenario, multiple access points (APs) function as O-RUs to maintain connectivity with transmitters, while the O-CUs and O-DUs are co-located within edge clouds (ECs) that manage connectivity with these APs.

For \textit{computation infrastructure}, the O-RAN architecture includes the Radio Access Network Intelligent Controller (RIC), Service and Orchestration Management (SMO), xAPPs, and rAPPs for software-based radio optimization and resource management of the network. The RIC is divided into near-real-time RIC (nRT-RIC) and non-real-time RIC (non-RT RIC) units based on network control loop execution times. The nRT-RIC hosts xAPPs that manage tasks such as transmitter reconfiguration, load balancing, interference detection, and mitigation utilizing machine learning models, while the non-RT RIC operates within the SMO, overseeing life-cycle management and configurations for all network elements. Additionally, the non-RT RIC hosts rAPPs, which handle higher-level tasks such as providing policy-based guidance to the xAPPs and coordinating xAPPs for optimized RAN management.
To fully leverage these functionalities, in our scenario, each EC integrates nRT-RIC and xAPPs, while the regional cloud (RC) houses the SMO, non-RT RIC, and rAPPs, as shown in Figure 1.

\begin{figure} [t]
\centerline{\includegraphics[width=1\columnwidth]{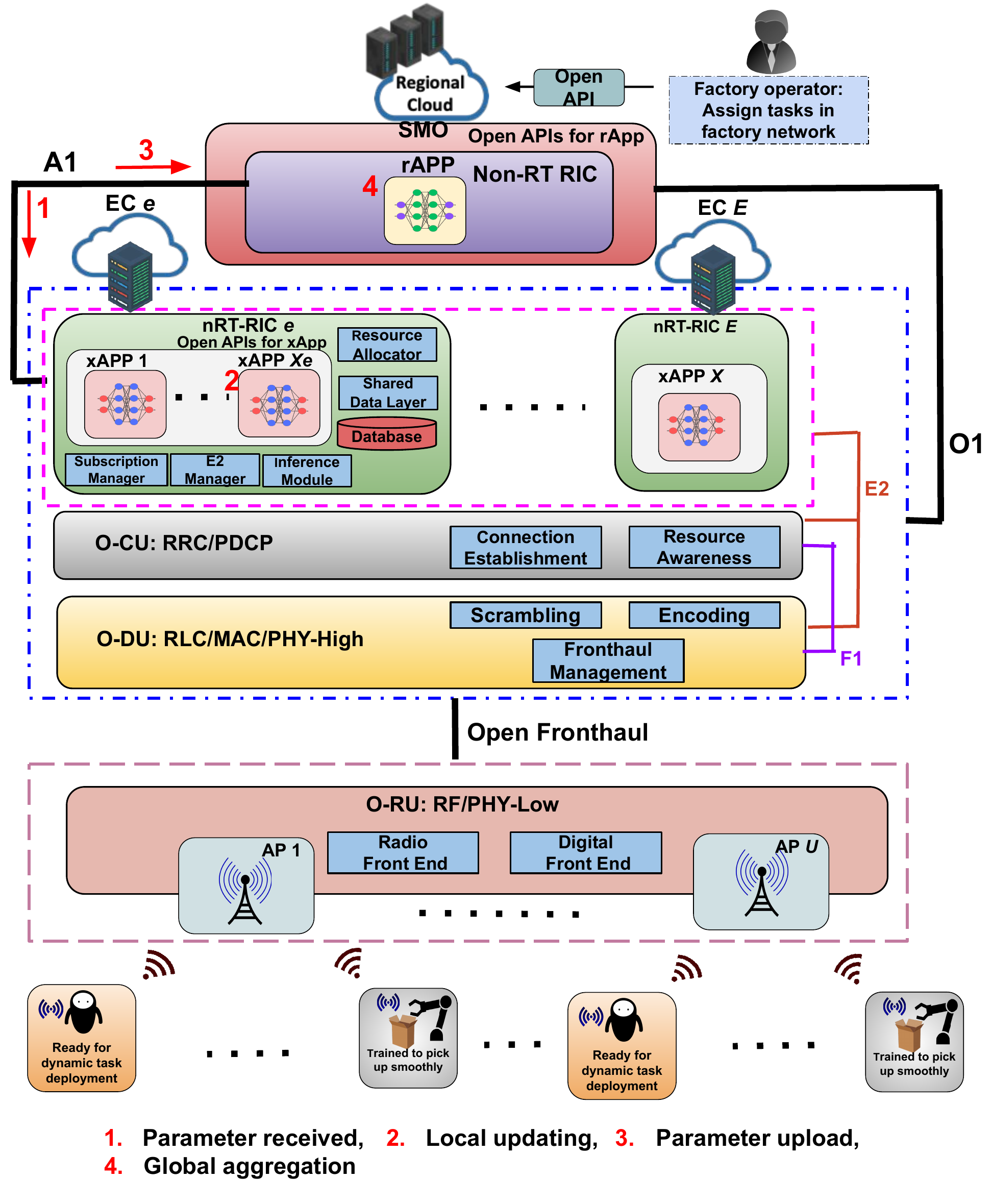}}
   \caption{The overall process of FedDRL framework in the proposed hierarchical O-RAN enabled zero-touch network platform.}
\end{figure}
\section{Network Model, Signaling Interface, and Problem Formulation}
\subsection{Network Model}
As illustrated in Figure 1, the proposed platform consists of a single RC, denoted as \( R \), along with \( E \) ECs, represented by the set \( \mathsf{E} = \{1,2,\dots,e,\dots,E\} \). Additionally, the platform includes \( U \) APs, indexed by the set \( \mathsf{U} = \{1,2,\dots,U\} \), \( N \) transmitters, collectively defined as \( \mathsf{N} = \{1,2,\dots,N\} \), and \( X \) xApps, represented by the set \( \mathsf{X} = \{1,2,\dots,X\} \).

It is assumed that each EC \( e \) serves a subset of APs, denoted as \( \mathsf{U}_{e} \), such that \( \mathsf{U}_{e} \subset \mathsf{U} \). Correspondingly, a subset of transmitters, \( \mathsf{N}_{e} = \{1,2,\dots,N_{e}\} \subset \mathsf{N} \), is covered by these \( \mathsf{U}_{e} \) APs. Furthermore, let \( \mathsf{N}_u = \{1,2,\dots,n,\dots,N_{u}\} \) represent the subset of transmitters associated with a specific AP \( u \), where \( u \in \mathsf{U}_{e} \), ensuring that \( \mathsf{N}_u \subset \mathsf{N}_{e} \). The transmitters within each \( \mathsf{N}_u \) are assumed to be uniformly distributed within the coverage area of their corresponding AP $u$.
Let, $\mathsf{M} = \{1,2,\dots,m,\dots,M\}$ and $\mathsf{P} = \{P_{1},P_{2},\dots,P_{k},\dots,P_{K}\}$ be the sets of MCS indices and available discrete transmit power levels, respectively. 

In the considered scenario, the multirobot system functions as a set of transmitters, with each robot's operational parameter reconfiguration decisions determined by xApps deployed on the nRT RICs. 
Specifically, within each nRT RIC, a subset of xApps denoted as \( \mathsf{X}_e \subset \mathsf{X} \), is responsible for managing the parameter reconfiguration decisions of the \( N_{e} \) associated transmitters, and $\mathsf{X}_e = \mathsf{N}_e$.  This direct mapping enables a one-to-one correspondence between xApps and transmitters, facilitating efficient and adaptive parameter reconfiguration processes within the network. The time axis is partitioned into discrete time steps, each of duration $\tau$. Moreover, the system employs orthogonal frequency division multiple access for communication between the APs and the transmitters.

\subsection{Signaling Procedure}
At the beginning of each time step \( t \), the xApps in each nRT-RIC determine the parameter reconfiguration decisions for transmitters based on key performance indicators (KPIs) and metrics. These include throughput, transmitter measurement reports such as signal-to-interference-plus-noise ratio (SINR), and metrics such as received power levels at the O-RUs from transmitters, all of which are gathered in the nRT-RICs during the previous time step \( t-1 \). 
\subsubsection{KPIs and Metrics Collection}
 The O-RUs collect transmitter measurement reports and uplink data through the physical uplink control channels and the physical uplink shared channels, respectively, while also measuring the uplink throughput and received power levels based on the uplink data. These measurements are subsequently reported to the O-DUs via the open fronthaul interface. The O-DUs then forward the collected reports to the O-CUs over the F1 interface. Finally, the O-CUs aggregate the reported measurements, such as SINR, throughput, and other relevant metrics before transmitting them to their respective nRT-RICs over E2 interface.
\subsubsection{Downlink Signaling}
Once the xApps in each nRT-RIC determine the parameter reconfiguration decisions based on the collected KPIs and metrics, the signaling process ensures the efficient delivery of these decisions to the transmitters. Initially, the nRT-RICs relay the computed reconfiguration decisions to the O-CUs via the E2 interface. The O-CUs then process these decisions and transmit the corresponding parameter reconfiguration commands to the O-DUs over the F1 interface. The O-DUs, in turn, forward the parameter reconfiguration instructions to the O-RUs through the open fronthaul interface. Finally, the O-RUs transmit the reconfiguration control signals to the respective transmitters over the physical downlink control channels. 

The proposed signaling procedure is detailedly illustrated in Figure 2.
\begin{figure} [t]
\centerline{\includegraphics[width=1\columnwidth]{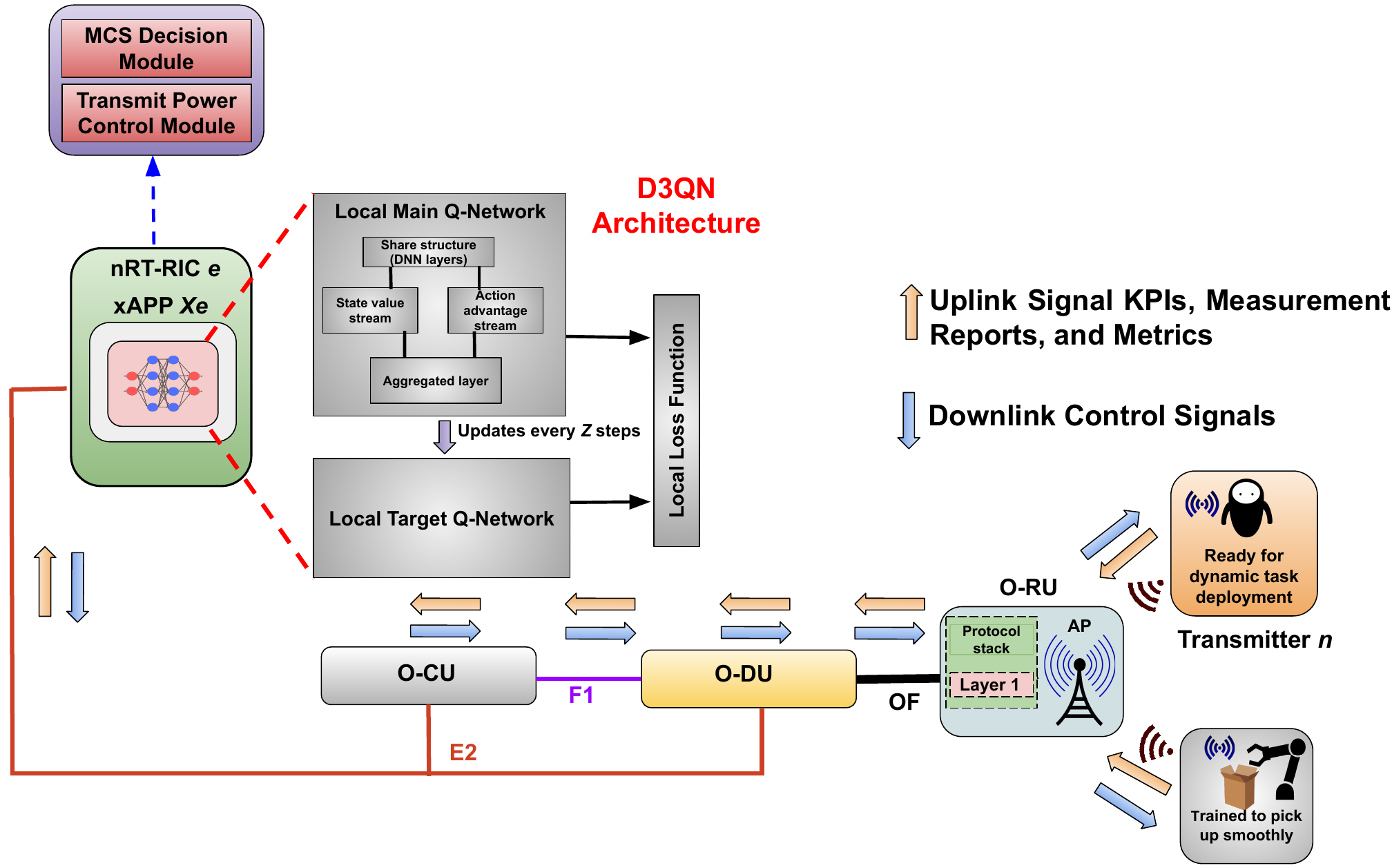}}
   \caption{Uplink KPIs, measurement reports, and metrics collection and downlink control signaling.}
\end{figure}

\subsection{Problem Formulation}
Due to the mobility of the transmitters, the wireless conditions of each communication link between the transmitters and the APs exhibit significant time variability. This dynamic nature necessitates the selection of an appropriate MCS index based on real-time channel conditions to optimize system throughput. However, while selecting a higher MCS index enables more data bits to be packed into transport blocks for transmission, it also increases the likelihood of block error rates and requires higher transmission power levels. 
Since the transmitters are battery-powered, increased transmission power levels lead to higher energy consumption and contribute to elevated interference levels, ultimately degrading overall QoS. 

To address these challenges, a joint optimization objective is formulated to maximize both throughput and transmission energy efficiency for each transmitter. To achieve this, xAPPs at nRT-RICs must solve the following problem.\\
\textbf{Optimization 1.} \textit{For the transmitters across the time steps from $t = 1$ to $t = T$, the selection of appropriate MCS indices and transmission power levels is framed as a joint optimization problem, formulated as:}
\vspace{-8.0pt}
\begin{equation}
\max_{\substack{m \in {\mathsf{M}} , P_{k} \in {\mathsf{P}}}} \sum_{t=1}^{T} \sum_{e \in \mathsf{E}} \sum_{u \in \mathsf{U_{e}}} \sum_{n \in \mathsf{N_{u}}} (\alpha_{1}\mathcal{T}_{u,n}^{e}(t) + \alpha_{2}\Gamma_{u,n}^{e}(t)),
\label{eq1}
\end{equation}
\text{subject to:}
\vspace{-5.0pt}
\begin{align}
\textbf{C1}: {\mathcal{T}}_{u,n}^{e}(t) \geq {\mathcal{T}}_{min}, \quad \ \ \forall e \in \mathsf{E}, \forall u \in \mathsf{U_{e}}, \forall n \in \mathsf{N_{u}}, \\
\textbf{C2}: P_{u,n}^{e}(t) \leq P_{max}, \quad \ \forall e \in \mathsf{E}, \forall u \in \mathsf{U_{e}}, \forall n \in \mathsf{N_{u}}, \\
\textbf{C3}: \Psi_{u,n}^{e}(t) \geq \Psi_{min}, \quad \  \forall e \in \mathsf{E}, \forall u \in \mathsf{U_{e}}, \forall n \in \mathsf{N_{u}}, \\
\textbf{C4}: \zeta_{u,n}^{e}(t) \leq \zeta_{max}, \quad \ \ \forall e \in \mathsf{E}, \forall u \in \mathsf{U_{e}}, \forall n \in \mathsf{N_{u}},
\end{align}
here, $\mathcal{T}_{u,n}^{e}(t)$ denotes the throughput of transmitter $n$ at time step $t$. Moreover, $\Gamma_{u,n}^{e}(t) = \frac{\beta^{m}_{n}(t)}{P_{u,n}^{e}(t)}$ represents the transmission energy efficiency, where, $\beta^{m}_{n}(t)$ corresponds to the data bits associated with the MCS indices for transmitter $n$ at time step $t$ and $P_{u,n}^{e}(t)$ denotes the transmit power level of the same transmitter at $t$. The terms $\alpha_{1}$ and $\alpha_{2}$ are weight factors, ranging from 0 to 1. 

The constraint \textbf{C1} ensures that the achieved throughput meets the minimum required level, \( \mathcal{T}_{min} \). Constraint \textbf{C2} enforces the maximum allowable transmit power, \( P_{max} \), to limit energy consumption and interference. Constraint \textbf{C3} guarantees that the reported SINR remains above the required threshold, $\Psi_{min}$, ensuring reliable communication. Finally, constraint \textbf{C4} limits the number of allocated physical resource blocks (PRBs) to \( \zeta_{max} \).

\section{Implementation of Intelligent Transmitter Parameter Reconfiguration}
\subsection{Description of the FedDRL-based Transmitter Reconfiguration Scheme}
Since \textbf{Optimization 1} is a joint optimization problem aimed at maximizing long-term throughput and transmission energy efficiency, it is difficult for conventional optimization techniques to solve due to the problem's complexity and non-convexity. Therefore, we reformulate \textbf{Optimization 1} as a Markov Decision Process (MDP) and employ D3QN \cite{10123947} to solve it. The formulated MDP can be represented as $\mathcal{M} = \mathcal{\{S, A, R, P\}}$, which includes the state space $\mathcal{S}$, the action space $\mathcal{A}$, the reward function $\mathcal{R}$, and the state transition probability $\mathcal{P}$. In practice, state transition probabilities are often unavailable, and D3QN addresses this by estimating these probabilities through interactions with the environment.
In D3QN, Deep Neural Networks (DNNs) are employed to approximate Q-values, and gradient descent, such as MGD is used to update the DNN parameters. To further mitigate reward bias, D3QN computes the final approximated Q-values by combining the value function ${V}(\mathcal{S}; \Theta^{{V}})$ for the state $\mathcal{S}$ with the advantage function ${A}(\mathcal{S}, \mathcal{A}; \Theta^{{A}})$ for each action $\mathcal{A}$. In practice, to ensure the uniqueness of both ${V}(\mathcal{S})$ and ${A}(\mathcal{S}, \mathcal{A})$, D3QN adjusts the Q-values by subtracting the average of all advantage values ${A}(\mathcal{S}, \mathcal{A}; \Theta^{{A}})$, as expressed by: $Q(\mathcal{S}, \mathcal{A}; \Theta) = {V(\mathcal{S}; \Theta^{{V}}}) + \left( {A(\mathcal{S}, \mathcal{A}}; \Theta^{{A}}) - \frac{1}{|\mathcal{A}^{*}|} \sum_{\mathcal{A}'} A(\mathcal{S}', \mathcal{A}'; \Theta) \right)$, which ensures the removal of any excess degrees of freedom.
Additionally, to enhance the stability of the learning process, D3QN maintains both a main DQN and a target DQN with similar structures. The target DQN’s parameters are periodically updated from the main DQN every $Z$ steps, allowing for more stable and reliable evaluation of action values over time, thereby enhancing the overall learning stability and performance. 

Moreover, to accelerate the convergence further, MGD \cite{9003425} is applied resulting in the following update rules:
\begin{equation}
\omega(t) = \eta \omega(t-1) + \nabla L(\Theta(t-1)),
\end{equation}
\begin{equation}
\Theta(t) = \Theta(t-1) - \alpha \omega(t),
\end{equation}
where \( \omega(t) \) represents the momentum term with \( \eta \) denoting the momentum attenuation factor and \( \nabla L(\Theta(t-1)) \) corresponding to the gradient of the loss function. Additionally, \( \Theta(t) \) denotes the model parameters at time step \( t \), while \( \alpha \) represents the learning rate.
The loss function $L(\Theta)$ is computed as the mean squared TD error over training batch:
\begin{equation}
L(\Theta) \triangleq \mathbb{E}\left[ (y^{D3QN} - Q_{\pi}(\mathcal{S}, \mathcal{A}; \Theta))^2 \right],
\end{equation}
where, $y^{D3QN} = \mathcal{R} + \gamma Q_{\pi}\big(\mathcal{S}',\underset{\mathcal{A'}}{\mathrm{argmax}}\ Q_{\pi}(\mathcal{S}',\mathcal{A}'; \Theta);\Theta^{-})$ with $\gamma$ being the discount factor, and $\Theta^{-}$ being the model parameters of the target DQN. The memory replay mechanism plays a critical role in D3QN by storing interaction experiences for model training. In this study, the PER mechanism \cite{schaul2015prioritized} is adopted instead of random sampling mechanism to enhance the efficiency of experience sampling and improve D3QN's learning performance.

For each transmitter $n$ in the multirobot system, there is a corresponding xAPP that maintains a local copy of the global D3QN model and operates as agent $n$. In this context, the global D3QN model refers to the model trained using the FedDRL approach. These independent agents leverage the global D3QN model during the inference phase to determine the optimal parameter configurations for their respective transmitters. Furthermore, in the FedDRL framework, the conventional elements of FL such as features, labels, and samples are replaced by state space, action space, and experience records, respectively. These elements are then utilized within the global D3QN model to facilitate decentralized decision making while preserving model generalization across distributed xApps. The detailed development of the state, action, and reward function of each agent $n$ is presented below:\\
\textit{\textbf{State:} The state for agent $n$ at time step $t$ can be defined as:}
\begin{equation}
\begin{split}
\small
\mathcal{S}_{{u, n}}^{e}(t) = \{\Psi_{u,n}^{e}(t-1),\mathcal{T}_{u,n}^{e}(t-1),\Upsilon_{u,n}^{e}(t-1),\\ \Phi_{u,n}^{e}(t-1), \mathcal{A}_{{u, n}}^{e}(t-1) \}, \quad \forall e \in \mathsf{E}, \forall u \in \mathsf{U_{e}}, \forall n \in \mathsf{N_{u}},
\end{split}
\end{equation}
where, $\mathcal{A}_{{u, n}}^{e}(t-1)$, $\Upsilon_{u,n}^{e}(t-1)$, and $\Phi_{u,n}^{e}(t-1)$ denote the action taken, the observed outcome of the taken action, and the received power level at time step $t-1$, respectively. Specifically, $\Upsilon_{u,n}^{e}(t-1)$ signifies whether $\mathcal{A}_{{u, n}}^{e}(t-1)$ was successful or not at time step $t-1$; if successful, a value of 1 is assigned, otherwise, a value of 0 is given.\\
\textit{\textbf{Action:} At time step $t$, the action for agent $n$ includes the joint selection of the MCS index and the transmission power level, defined as follows:}
\begin{equation}
\small
\mathcal{A}_{{u, n}}^{e}(t) = \{m \in \mathsf{M}, P_{k}\in \mathsf{P} \}, \quad \forall e \in \mathsf{E}, \forall u \in \mathsf{U_{e}}, \forall n \in \mathsf{N_{u}},
\end{equation}
\textit{\textbf{Reward Function:} At time step $t$, each agent $n$ independently takes an action and subsequently receives feedback in the form of a \textbf{global reward} from the environment and computed as: }
\begin{equation}
\small  
\mathcal{R}^{global}(t) = \lambda \sum_{e \in \mathsf{E}} \sum_{u \in \mathsf{U_{e}}} \sum_{n \in \mathsf{N_{u}}} R_{u,n}^{e}(t), \forall e \in \mathsf{E}, \forall u \in \mathsf{U_{e}}, \forall n \in \mathsf{N_{u}},
\end{equation}
where, $\lambda$ = $\frac{1}{\sum_{e \in \mathsf{E}} \sum_{u \in \mathsf{U_{e}}} \vert \mathsf{N_{u}} \vert}$ and  $R_{u,n}^{e}(t)$ can be expressed as:
\begin{equation}
R_{u,n}^{e}(t) = \begin{cases} 
 \alpha_{1} \mathcal{T}_{u,n}^{e}(t) + \alpha_{2} \Gamma_{u,n}^{e}(t) + \tau_{1} \Omega(t), & \text{if $\mathcal{A}_{{u, n}}^{e}(t)$} \\ 
 & \text{successful} \\
 -\alpha_{1} \mathcal{T}_{u,n}^{e}(t) - \tau_{2} C - \tau_{3} C, & \text{if $\mathcal{A}_{{u, n}}^{e}(t)$} \\
 & \text{failed}  
\end{cases}
\end{equation}
where, $C $ is a constant and $\tau_{1}, \tau_{2}$, and $\tau_{3}$ are the weight factors, ranges from 0 to 1. Additionally, $\Omega (t)$ = $\Upsilon_{u,n}^{e}(t)$.

\begin{algorithm}
\caption{FedDRL-based Transmitter Reconfiguration.}
\begin{algorithmic}[1]
\footnotesize
\State \textbf{Initialize:} Global momentum and D3QN parameters for each agent $n$
\For{$r$ $= 1$ to $R$} 
\For{each agent $n$ in parallel}, $\forall e \in \mathsf{E}, \forall u \in \mathsf{U_{e}}, \forall n \in \mathsf{N_{u}}$ 
\State Receive global momentum parameters $\omega^{global}$ and D3QN parameters $\Theta^{global}$;
\EndFor
\State Initialize states $\mathcal{S}_{{u, 1}}^{e}, \mathcal{S}_{{u, 2}}^{e}, \ldots, \mathcal{S}_{{u, n}}^{e}$; $\forall e \in \mathsf{E}, \forall u \in \mathsf{U_{e}}, \forall n \in \mathsf{N_{u}}$
\For{$t$ $= 1$ to $T$}
\For{each agent $n$}
\State Select an action utilizing decaying \( \epsilon \)-greedy policy
\EndFor
\State Execute joint action $\mathcal{A}(t) = (\mathcal{A}_{{u, 1}}^{e}(t), \mathcal{A}_{{u, 2}}^{e}(t), \ldots, \mathcal{A}_{{u, n}}^{e}(t))$;  
\State Observe global reward $\mathcal{R}^{global}(t)$, $ \forall e \in \mathsf{E}, \forall u \in \mathsf{U_{e}}, \forall n \in \mathsf{N_{u}}$;
\State Observe next states $\mathcal{S}_{{u, 1}}^{e}(t+1), \mathcal{S}_{{u, 2}}^{e}(t+1), \ldots, \mathcal{S}_{{u, n}}^{e}(t+1)$;
\State Store transition in local buffer $\mathcal{D}_{u,1}^{e}$, $\mathcal{D}_{u,2}^{e}$, \ldots, $\mathcal{D}_{u,n}^{e}$;
            \For{each agent $n$}
                \If {over every $\varkappa$ time steps}
                \State Sample a minibatch of experiences from $\mathcal{D}_{u,n}^{e}$ using PER;
                \For{each experience in the minibatch}
                    \State Compute target Q-value, $Q_{u,n}^{e, target}$
                    \State Compute predicted Q-value, $Q_{u,n}^{e, pred}$
                    \State Compute loss, $L_{u,n}^{e}(\Theta(t))$
                \EndFor
                \State Update local momentum parameters using Eq. (6);
               \State Update local model parameters using Eq. (7);
                \EndIf
               \State Update local states: $\mathcal{S}_{{u, n}}^{e}(t) = \mathcal{S}_{{u, n}}^{e}(t+1)$;
            \EndFor
        \EndFor
\State Send locally updated momentum parameters ($\omega_{u,1}^{e}, \omega_{u,2}^{e}, \dots, \omega_{u,n}^{e}$) and model parameters ($\Theta_{u,1}^{e}, \Theta_{u,2}^{e}, \dots, \Theta_{u,n}^{e}$) to the non-RT-RIC;
\State Aggregate and update global momentum and model parameters using Eq. (13) and Eq. (14), respectively at the non-RT-RIC; 
\EndFor
\end{algorithmic}
\end{algorithm}

In FedDRL, the primary goal of federating is to construct a global D3QN model that maximizes long-term system throughput and transmission efficiency, while accelerating environmental exploration through the sharing of experiences. 
The FedDRL process can be divided into the below phases:\\
\textit{\textbf{Parameter Receive Phase:}} At global round $r \in R$, agents operating within each nRT-RIC $e$ obtain the most recent global momentum and model parameters from the non-RT-RIC through the A1 interface. Upon receipt, these agents replace their existing local momentum and model parameters with the newly received global parameters.\\
\textit{\textbf{Inference and Local Parameter Update Phase:}} 
After obtaining the new local momentum and model parameters, each agent $n$ immediately initiates real-time inference, continuing this process at each subsequent time step. Furthermore, over the course of $T$ time steps within global round $r$, agents periodically compute gradients after every $\varkappa$ time steps, leveraging their respective local experiences. These experiences are collected as minibatches $\delta$ using the PER mechanism from the agent's local buffer in the shared database. The gradients are then used to incrementally update the local momentum and model parameters according to (6) and Eq. (7), respectively.\\
\textit{\textbf{Parameter Upload Phase:}} Upon completing $T$ time steps, each agent $n$ transmits the locally updated momentum and model parameters to the non-RT-RIC through the A1 interface.\\
\textit{\textbf{Global Aggregation and Feedback Phase:}} Upon receiving all the local momentum and model parameters from agents, the rAPP at non-RT-RIC aggregates these inputs by utilizing FedAvg algorithm \cite{FedAvg}. This process results in the updated global momentum and model parameters, as detailed below:
\begin{equation}
\omega^{global} = \frac{\sum_{e \in \mathsf{E}} \sum_{u \in \mathsf{U}_{e}} \sum_{n \in \mathsf{N_{u}}} \omega_{u, n}^{e}}{\sum_{e \in \mathsf{E}} \sum_{u \in \mathsf{U_{e}}} \vert \mathsf{N_{u}} \vert}, 
\end{equation}
\begin{equation}
\Theta^{global} = \frac{\sum_{e \in \mathsf{E}} \sum_{u \in \mathsf{U}_{e}} \sum_{n \in \mathsf{N_{u}}} \Theta_{u,n}^{e}}{\sum_{e \in \mathsf{E}} \sum_{u \in \mathsf{U_{e}}} \vert \mathsf{N_{u}} \vert}.
\end{equation}
Finally, at global round $r+1$, the non-RT-RIC transmits back the updated global momentum $\omega^{global}$ and the model $\Theta^{global}$ parameters to all agents. The whole process is depicted in Figure 1 and summarized in Algorithm 1.
\subsection{Complexity Analysis}
Next, the computational complexity of the proposed scheme is analyzed. Each transmitter $n$ employs a local D3QN model implemented as a fully connected feedforward neural network encompassing $\omega$ hidden layers, with each hidden layer $f$ containing $\phi_f$ neurons. To quantify the complexity associated with updating the parameters of these fully connected layers, the function $\Phi(\cdot)$ is introduced. Moreover, due to the dueling architecture, the output layer is partitioned into two separate streams: one for the state-value function and another for the advantage function. Thus, considering $R$ global training rounds, $T$ time steps per round, $N$ transmitters, and a mini-batch size of $\delta$, the computational complexity of the FedDRL can be formulated as:
\begin{equation}  
\footnotesize
 \mathcal{O}(RTN\delta\Lambda\sigma) = \mathcal{O} \left (RTN\delta\Lambda\Phi \left( \sum_{f=1}^{\omega-1} \phi_f \phi_{f+1} + \phi_{\omega} + a\phi_{\omega} \right)\right),  
\end{equation}  
where, $\Lambda = \frac{R}{\rho}$ with $\rho$ representing the total number of model parameter aggregations occurring within $R$. Moreover, $\sigma$ denotes the function $\Phi(\cdot)$, and $a$ corresponds to the dimensionality of the $\mathcal{A}_{{u, n}}^{e}$, $\forall e \in \mathsf{E}, \forall u \in \mathsf{U_{e}}, \forall n \in \mathsf{N_{u}}$. 
\section{Performance Evaluation}
This section outlines the system and algorithm parameters and evaluates the performance of FedDRL in terms of convergence, adaptiveness, and robustness. The experiments are conducted using a Python 3.9 simulation platform, with the TensorFlow-Keras library utilized to construct the DNNs.

\subsection{Simulation Environment Setup}
 The simulation scenario involves a single RC, a single EC $e$, $U_{e}$ APs, and $N_{e}$ transmitters, where, $U_{e} = 4$ and $N_{e} = [12, 20]$, unless stated otherwise. The coverage area of each AP is set to 100m\cite{9272626}. The transmitters move randomly at a speed of 3 km/h, simulating real-world conditions. 
 The channel model adopted from \cite{9272626} is the product of three components: random shadowing, random small-scale fading, and a distance-dependent path-loss model. The distance-dependent pathloss can be modeled as $P(d) =-\xi - 10\varphi\log_{10}(\frac{d}{d_0})$,  where, the path-loss exponent $\varphi = 3$ and the reference distance $d_{0}$ = 1m  \cite{9272626}.
 The number of available MCSs $M$ for each transmitter is set to 29, in accordance with the 3GPP standard\cite{3gppTS38214}. 
 The spectral efficiency (in bits/sec/Hz) of each MCS index for 5G New Radio physical uplink shared channel messages, which determines the corresponding data bits can be referenced from \cite{3gppTS38214}.   
 At each time step, each transmitter can be allocated a fix number of PRBs from the set $\mathsf{C} = \{1,2,3,\dots,\zeta_{n}\}$ for uplink transmission. Table I summarizes the network simulation parameters.

 \begin{table}[!h]
\centering
\caption{Network system parameter setups.}
\footnotesize 
\begin{tabularx}{\columnwidth}{Xcc}
\toprule
\textbf{Network Parameters} & \textbf{Values} \\
\midrule
Carrier frequency & 3.0\,GHz \\
PRBs bandwidth & 180\,KHz \\
Noise power & -110\,dBm \\
Transmit power levels & \{-8.4, -2.3, 0, 4, 7, 9\}\,dBm \\
Received SINR margin & [-6.7, 11.7]\,dB \\
\bottomrule
\end{tabularx}
\end{table} 

Furthermore, simulations are executed on a 2021 MacBook Pro, equipped with an Apple M1 Pro chip, featuring a 64-bit operating system, an 8-core CPU, a 14-core GPU, and 16GB of unified RAM. Table II details the D3QN configurations used in the simulation.

 \begin{table}[!h]
\centering
\caption{D3QN parameters for simulation.}
\footnotesize 
\begin{tabularx}{\columnwidth}{Xcc}
\toprule
\textbf{D3QN Parameters} & \textbf{Values} \\
\midrule
PER memory size & 4000\\ 
NN-layers (common)& 32$\times$32 \\
NN-layers (value stream)& 32$\times$1 \\
NN-layers (advantage stream)& 32$\times$174 \\
Activation & tanh\\
Local Optimizer & MGD\\ 
Local learning rate ($\alpha$) & 0.001 \\
Discount factor ($\gamma$)  & 0.995 \\
Initial ($\epsilon$), decay rate, minimum ($\epsilon$) & 1.0, 0.9995, 0.1 \\
Local minibatch size ($\delta$)  & 32 \\ 
Target DQN updating period ($Z$)  & 200 \\ 
Number of global rounds ($R$)  & 60\\ 
Number of time steps per global round ($T$) & 500\\ 
Number of time steps for gradients update ($\varkappa$) & 50\\ 
\bottomrule
\end{tabularx}
\end{table}

 To validate the efficacy of the proposed framework, the following two methods are simulated for comparison:\\
 \textit{\textbf{Independent DRL (IDRL):}} Each agent independently trains its local model and momentum parameters without utilizing a global model or shared momentum parameters.\\
 \textit{\textbf{Random Actions (RA):}} Agents select MCS indices and transmission power levels in a random manner.
 
 In addition, four performance metrics are considered: (1) system throughput, (2) normalized cumulative rewards, (3) average transmission energy consumption, and (4) normalized average transmission energy efficiency. 
\subsection{Simulation Results}

\begin{figure} [!b]
    \centering
  \subfloat[\label{1a}]{%
       \includegraphics[width=0.50\linewidth]{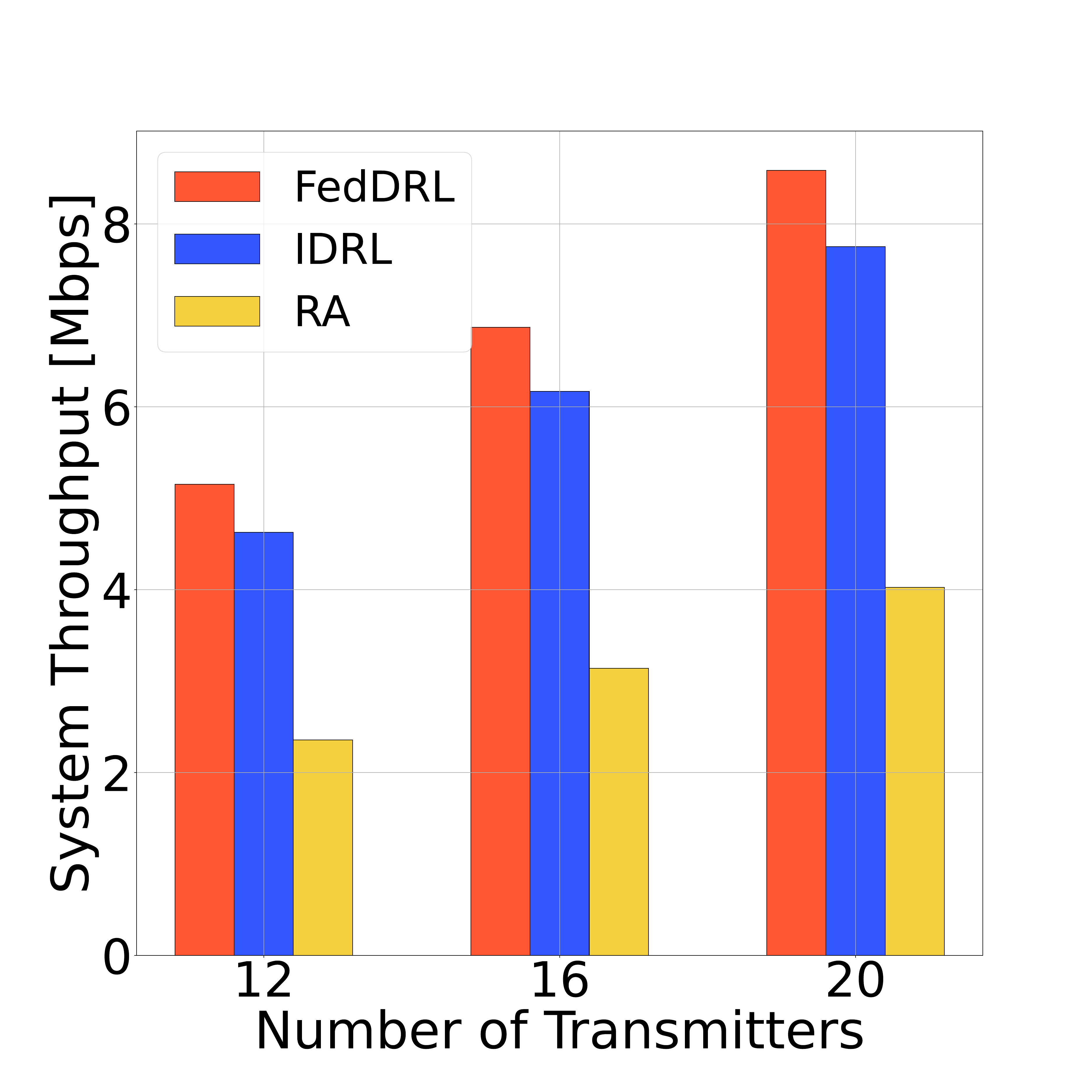}}
    \hfill
  \subfloat[\label{1b}]{%
        \includegraphics[width=0.50\linewidth]{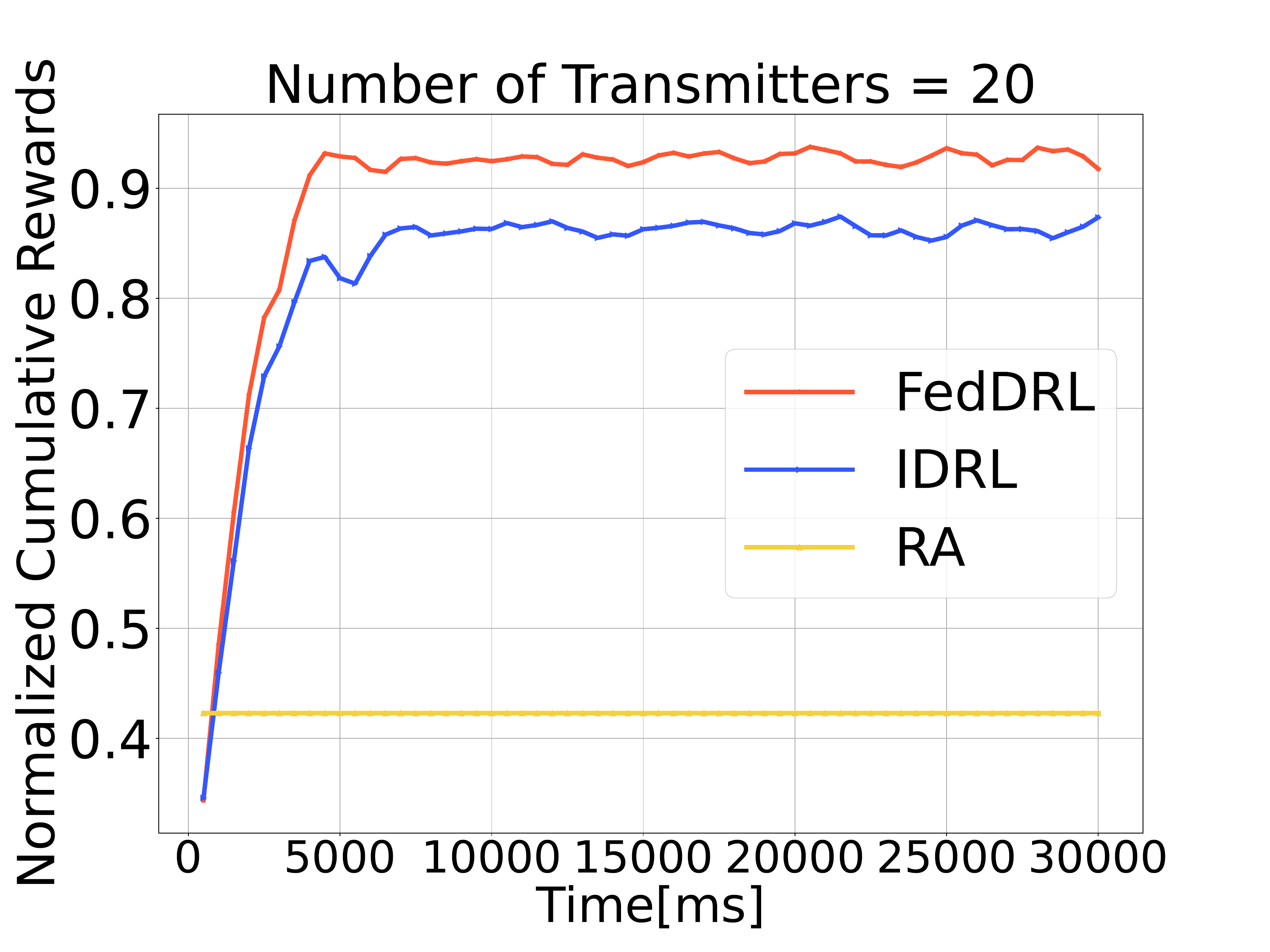}}
    \\
\vspace{-12pt}
  \subfloat[\label{1c}]{%
        \includegraphics[width=0.50\linewidth]{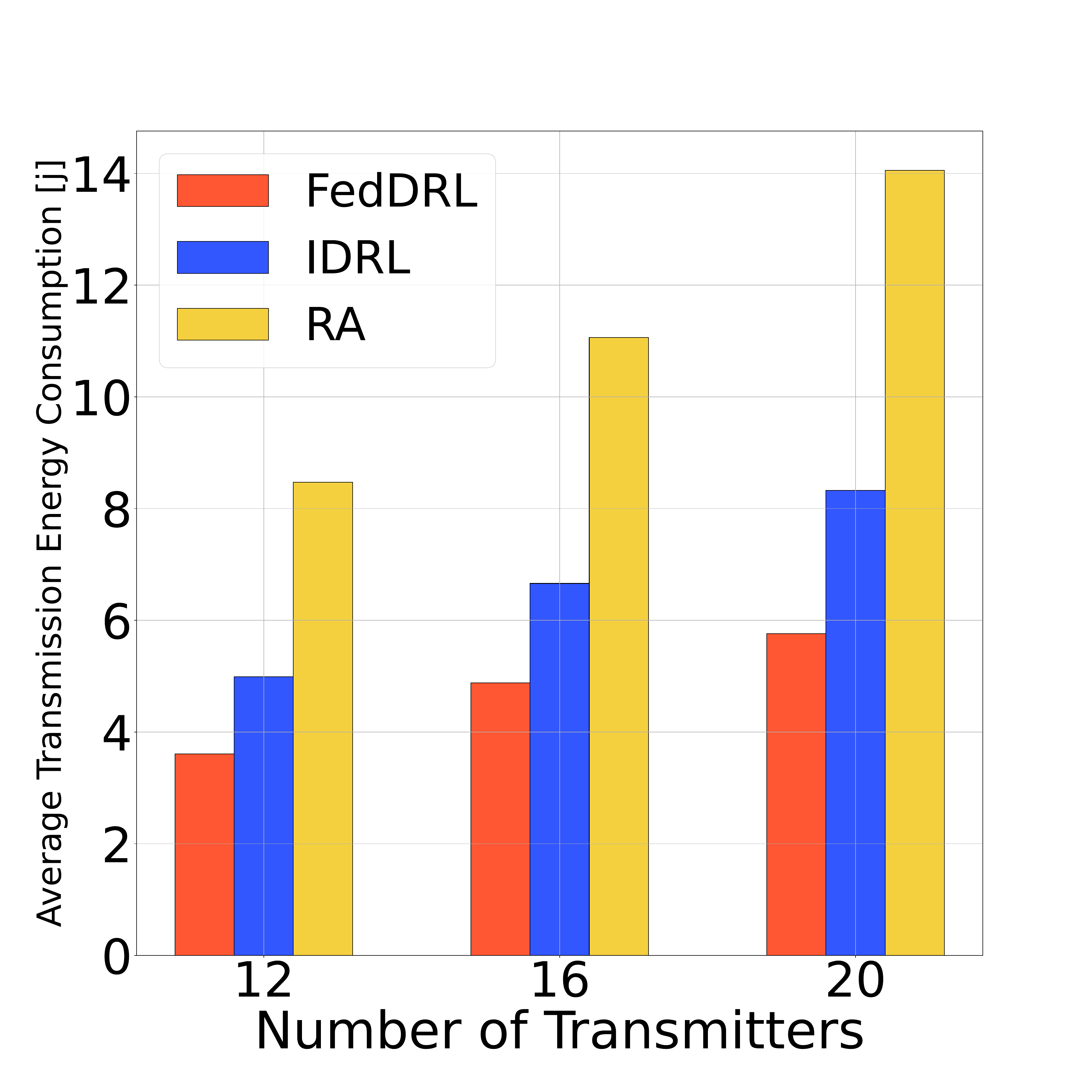}}
    \hfill
  \subfloat[\label{1d}]{%
        \includegraphics[width=0.50\linewidth]{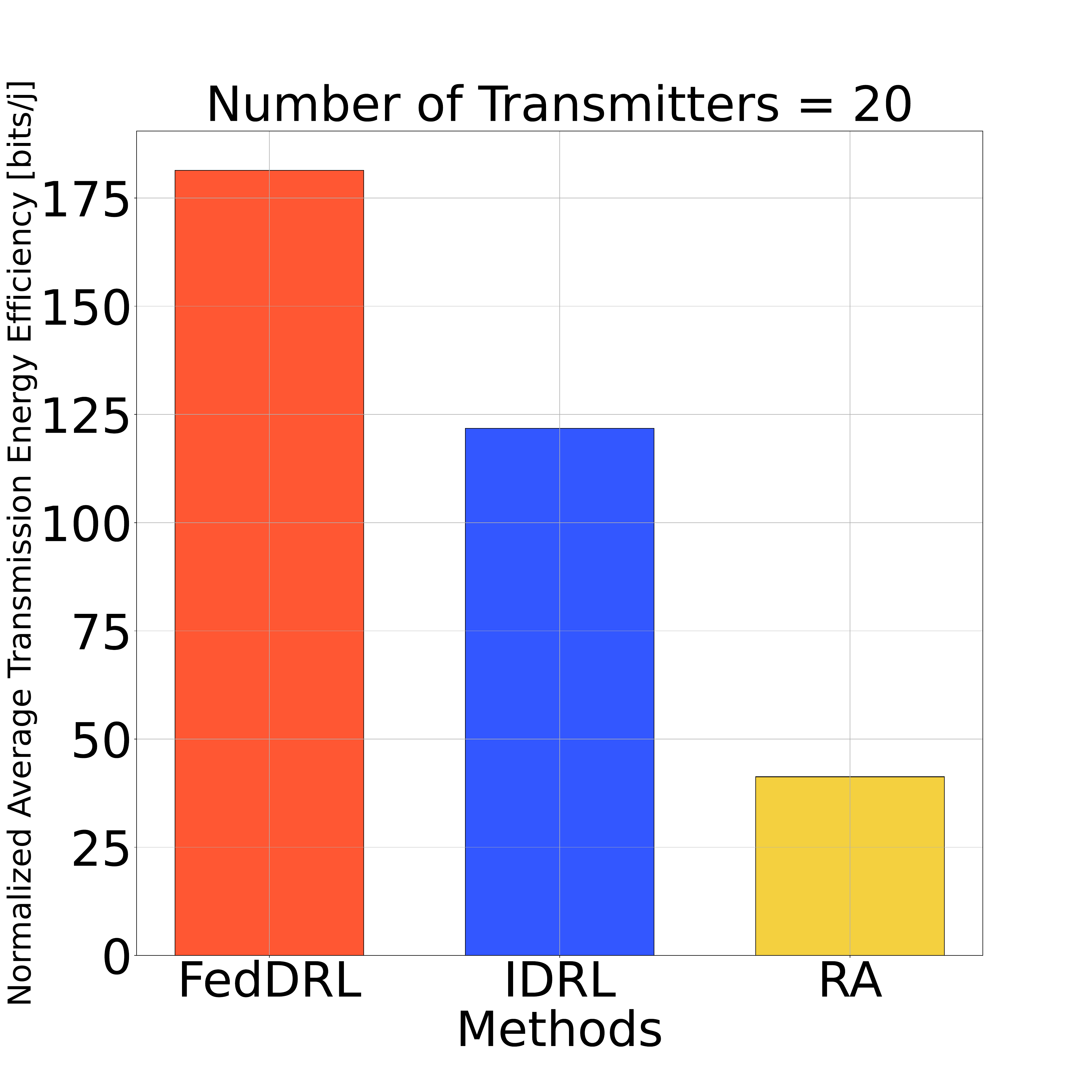}}

  \caption{(a) System throughput vs varying number of transmitters; (b) normalized cumulative rewards vs time; (c) average transmission energy consumption vs varying number of transmitters; (d) normalized average transmission efficiency for the proposed FedDRL and comparison methods.}
  \label{fig3} 
\end{figure}

Figure 3(a) presents the system throughput for varying numbers of transmitters, comparing the proposed FedDRL approach with the benchmark methods, namely IDRL and RA. The results indicate that FedDRL consistently outperforms both IDRL and RA in terms of system throughput. Specifically, when the number of transmitters is set to 20, FedDRL achieves throughput gains of 12\% and 53\% over IDRL and RA, respectively. This improvement is attributed to the ability of FedDRL to prioritize critical experiences, stabilize updates between distributed agents, and facilitate decentralized collaboration, collectively reducing variance in value updates and enhancing decision-making efficiency.

Figure 3(b) illustrates the normalized cumulative rewards over time steps for FedDRL, IDRL, and RA, with the number of transmitters fixed at 20. The results demonstrate that FedDRL achieves significantly higher rewards compared to IDRL. This performance gain can be explained by the same factors that contribute to the superior performance observed in Figure 3(a). Additionally, FedDRL exhibits faster convergence than IDRL due to its integration of global momentum, which optimizes the loss functions of individual agents more effectively.

Figure 3(c) compares the average transmission energy consumption between different numbers of transmitters for FedDRL, IDRL, and RA. The results reveal that FedDRL consistently achieves lower energy consumption than the benchmark methods. For example, when the number of transmitters is set to 20, FedDRL reduces transmission energy consumption by 28\% and 59\% compared to IDRL and RA, respectively. This efficiency gain aligns with the improvements discussed in Figure 3(a), strengthening the effectiveness of FedDRL in balancing transmission energy efficiency and performance.

Figure 3(d) illustrates the normalized average transmission efficiency for different numbers of transmitters. The results indicate that FedDRL achieves 32\% and 77\% higher transmission efficiency compared to IDRL and RA, respectively, when the number of transmitters is set to 20. The underlying factors driving this improvement are consistent with those discussed in Figure 3(a), which highlights the advantages of FedDRL in optimizing both throughput and energy efficiency. 



\section{Conclusions and Future Works}

This paper presents a zero-touch network platform designed for smart factory applications, integrating a hierarchical Open Radio Access Network architecture with federated deep reinforcement learning (FedDRL). The introduced platform enables intelligent parameter reconfiguration of multirobot systems by leveraging FedDRL-driven adaptive transmitter optimization, specifically through modulation and coding scheme selection and transmission power control. By federating learning across distributed agents, the platform enhances system throughput, transmission energy efficiency, and overall transmission performance in dynamic and heterogeneous factory environments. The research findings highlight the platform’s potential to advance Industry 4.0 objectives by ensuring seamless data-driven adaptation in complex industrial ecosystems.


Expanding beyond the current single-edge-cloud assumption, future research will investigate a multi-edge-cloud environment, explicitly addressing the mobility of transmitters across multiple near-real-time RICs (nRT-RICs). This extension will clarify how the proposed system adapts as transmitters transition between the control domains of different nRT-RICs, thus capturing the complexities introduced by dynamic wireless conditions and interactions among multiple access points of the different nRT-RICs. Additionally, real-world deployment and comprehensive testing in industrial settings will be emphasized to validate the practicality and scalability of the proposed platform under diverse operational scenarios.


\section*{Acknowledgement}
This work was supported in part by Cisco Research, the National Science Foundation (NSF) under Grant CNS-221034, Meta 2022 AI4AI Research, and the NC Space Grant. We appreciate Dr. Luis Tello-Oquendo's technical comments on helping with the production of this work.


\bibliographystyle{IEEEtran}
\bibliography{IEEEabrv,bib}

\begin{thebibliography}{10}
\providecommand{\url}[1]{#1}
\csname url@samestyle\endcsname
\providecommand{\newblock}{\relax}
\providecommand{\bibinfo}[2]{#2}
\providecommand{\BIBentrySTDinterwordspacing}{\spaceskip=0pt\relax}
\providecommand{\BIBentryALTinterwordstretchfactor}{4}
\providecommand{\BIBentryALTinterwordspacing}{\spaceskip=\fontdimen2\font plus
\BIBentryALTinterwordstretchfactor\fontdimen3\font minus \fontdimen4\font\relax}
\providecommand{\BIBforeignlanguage}[2]{{%
\expandafter\ifx\csname l@#1\endcsname\relax
\typeout{** WARNING: IEEEtran.bst: No hyphenation pattern has been}%
\typeout{** loaded for the language `#1'. Using the pattern for}%
\typeout{** the default language instead.}%
\else
\language=\csname l@#1\endcsname
\fi
#2}}
\providecommand{\BIBdecl}{\relax}
\BIBdecl

\bibitem{9272626}
K.-C. Chen, S.-C. Lin, J.-H. Hsiao, C.-H. Liu, A.~F. Molisch, and G.~P. Fettweis, ``Wireless networked multirobot systems in smart factories,'' \emph{Proceedings of the IEEE}, vol. 109, no.~4, pp. 468--494, 2021.

\bibitem{10004601}
I.~Ashraf, Y.~B. Zikria, S.~Garg, Y.~Park, G.~Kaddoum, and S.~Singh, ``Zero touch networks to realize virtualization: Opportunities, challenges, and future prospects,'' \emph{IEEE Network}, vol.~36, no.~6, pp. 251--259, 2022.

\bibitem{9913206}
E.~Coronado, R.~Behravesh, T.~Subramanya, A.~Fernàndez-Fernàndez, M.~S. Siddiqui, X.~Costa-Pérez, and R.~Riggio, ``Zero touch management: A survey of network automation solutions for 5g and 6g networks,'' \emph{IEEE Communications Surveys \& Tutorials}, vol.~24, no.~4, pp. 2535--2578, 2022.

\bibitem{10004596}
T.~Wang, J.~Li, W.~Wei, W.~Wang, and K.~Fang, ``Deep-learning-based weak electromagnetic intrusion detection method for zero touch networks on industrial iot,'' \emph{IEEE Network}, vol.~36, pp. 236--242, 2022.

\bibitem{o-ran2023}
{O-RAN Alliance}, ``{O-RAN Architecture Description 9.0},'' {O-RAN Alliance}, Technical Specification, September 2024.

\bibitem{9376232}
B.~Balasubramanian, E.~S. Daniels, M.~Hiltunen, R.~Jana, K.~Joshi, R.~Sivaraj, T.~X. Tran, and C.~Wang, ``Ric: A ran intelligent controller platform for ai-enabled cellular networks,'' \emph{IEEE Internet Computing}, vol.~25, no.~2, pp. 7--17, 2021.

\bibitem{10355063}
S.-Y. Lien, Y.-C. Huang, C.-C. Tseng, S.-C. Lin, C.-L. I, X.~Xu, and D.-J. Deng, ``Universal vertical application adaptation for o-ran: Low-latency ric and autonomous intelligent xapp generation,'' \emph{IEEE Communications Magazine}, vol.~62, no.~5, pp. 80--86, 2024.

\bibitem{Ahmed2025}
F.~Ahmed, M.~Lee, S.~Subramaniam, M.~Matsuura, H.~Hasegawa, and S.-C. Lin, ``Enhancing network traffic analysis in o-ran enabled next-generation networks through federated multi-task learning,'' in \emph{IEEE Wireless Communications and Networking Conference (WCNC)}, Milan, Italy, March 2025.

\bibitem{FedAvg}
B.~McMahan, E.~Moore, D.~Ramage, S.~Hampson, and B.~A.~y. Arcas, ``{Communication-Efficient Learning of Deep Networks from Decentralized Data},'' in \emph{Proceedings of the 20th International Conference on Artificial Intelligence and Statistics}.\hskip 1em plus 0.5em minus 0.4em\relax PMLR, 20--22 Apr 2017, pp. 1273--1282.

\bibitem{10138331}
S.-C. Lin, C.-H. Lin, and M.~Lee, ``Privacy-preserving serverless edge learning with decentralized small-scale mobile data,'' \emph{IEEE Network}, vol.~38, no.~2, pp. 264--271, 2024.

\bibitem{9631391}
A.~A. Al-Saedi, V.~Boeva, and E.~Casalicchio, ``Reducing communication overhead of federated learning through clustering analysis,'' in \emph{2021 IEEE Symposium on Computers and Communications (ISCC)}, 2021, pp. 1--7.

\bibitem{9771700}
A.~K. Singh and K.~Khoa~Nguyen, ``Joint selection of local trainers and resource allocation for federated learning in open ran intelligent controllers,'' in \emph{2022 IEEE Wireless Communications and Networking Conference (WCNC)}, 2022, pp. 1874--1879.

\bibitem{Asad2024}
M.~Asad and S.~Otoum, ``Federated learning for efficient spectrum allocation in open ran,'' \emph{Cluster Computing}, vol.~27, pp. 11\,237--11\,247, 2024.

\bibitem{10130620}
S.~Ben~Saad, B.~Brik, and A.~Ksentini, ``Toward securing federated learning against poisoning attacks in zero touch b5g networks,'' \emph{IEEE Transactions on Network and Service Management}, vol.~20, no.~2, pp. 1612--1624, 2023.

\bibitem{9838376}
S.~Roy, H.~Chergui, L.~Sanabria-Russo, and C.~Verikoukis, ``A cloud native sla-driven stochastic federated learning policy for 6g zero-touch network slicing,'' in \emph{ICC 2022 - IEEE International Conference on Communications}, 2022, pp. 4269--4274.

\bibitem{10012789}
H.~Erdol, X.~Wang, P.~Li, J.~D. Thomas, R.~Piechocki, G.~Oikonomou, R.~Inacio, A.~Ahmad, K.~Briggs, and S.~Kapoor, ``Federated meta-learning for traffic steering in o-ran,'' in \emph{2022 IEEE 96th Vehicular Technology Conference (VTC2022-Fall)}, 2022, pp. 1--7.

\bibitem{10416344}
Z.~A.~E. Houda, H.~Moudoud, and B.~Brik, ``Federated deep reinforcement learning for efficient jamming attack mitigation in o-ran,'' \emph{IEEE Transactions on Vehicular Technology}, vol.~73, no.~7, pp. 9334--9343, 2024.

\bibitem{10123947}
Y.~Ji, Y.~Wang, H.~Zhao, G.~Gui, H.~Gacanin, H.~Sari, and F.~Adachi, ``Multi-agent reinforcement learning resources allocation method using dueling double deep q-network in vehicular networks,'' \emph{IEEE Transactions on Vehicular Technology}, vol.~72, no.~10, pp. 13\,447--13\,460, 2023.

\bibitem{9003425}
W.~Liu, L.~Chen, Y.~Chen, and W.~Zhang, ``Accelerating federated learning via momentum gradient descent,'' \emph{IEEE Transactions on Parallel and Distributed Systems}, vol.~31, no.~8, pp. 1754--1766, 2020.

\bibitem{schaul2015prioritized}
T.~Schaul, J.~Quan, I.~Antonoglou, and D.~Silver, ``Prioritized experience replay,'' \emph{arXiv preprint arXiv:1511.05952}, 2015.

\bibitem{3gppTS38214}
3GPP, \emph{{{TS 38.214, NR; Physical layer procedures for data}}}, Sep 2023.

\end{thebibliography}
\end{document}